\documentclass[aps,showpacs,preprint,superscriptaddress]{revtex4}
\usepackage{psfig}
\newcommand{\bea}{\begin{eqnarray}}
\newcommand{\beq}{\begin{equation}}
\newcommand{\eea}{\end{eqnarray}}
\newcommand{\eeq}{\end{equation}}

\usepackage{epsf}
\usepackage[dvips]{graphicx}

\begin{document}
\bibliographystyle{revtex}

\begin{center}

{\bfseries RANDOM MATRIX THEORY AND ANALYSIS OF NUCLEUS-NUCLEUS
COLLISION AT HIGH ENERGIES}

\vskip 5mm

E.\ I.\ Shahaliev $^{1,2 \dag}$, R.\ G.\ Nazmitdinov $^{3,4}$, A.\
A.\ Kuznetsov $^{1}$, M.\ K.\ Suleymanov $^{1}$, O.\ V. Teryaev
$^{4}$


{\small
(1) {\it High Energy Physics Laboratory, Joint Institute
for Nuclear Research, 141980, Dubna, Russia}
\\
(2) {\it Institute of Radiation Problems, 370143, Baku,
Azerbaijan}
\\
(3) {\it Departament de F{\'\i}sica, Universitat de les Illes
Balears, E-07122 Palma de Mallorca, Spain}
\\
(4) {\it Bogoliubov Laboratory of Theoretical Physics, Joint
Institute for Nuclear Research, 141980 Dubna, Russia}

$\dag$ {\it E-mail: shah@sunhe.jinr.ru }}
\end{center}

\vskip 5mm

\begin{center}
\begin{minipage}{150mm}
\centerline{\bf Abstract}

We propose a novel method for analysis of experimental data
obtained at relativistic nucleus-nucleus collisions. The method,
based on the ideas of Random Matrix Theory, is applied to detect
systematic errors that occur at measurements of momentum
distributions of emitted particles. The unfolded momentum
distribution is well described by the Gaussian orthogonal ensemble
of random matrices, when the uncertainty in the momentum
distribution is maximal. The method is free from unwanted
background contributions.

\end{minipage}
\end{center}

\vskip 10mm

Relativistic heavy ion collisions are among major experimental
tools that allow to get insight into nuclear dynamics at high
excitation energies and large baryon densities. It is expected
that in central collisions, at energies that are and will be soon
available at SPS(CERN), RHIC(BNL) and LHC(CERN), the nuclear
density may exceed by tens times the density of stable nuclei. At
such extreme conditions one would expect that a final product of
heavy ion collisions could present a composite system that
consists of free nucleons, quarks and quark-gluon plasma.
However, identification of the quark-gluon plasma, for example, is
darken due to a multiplicity of secondary particles created at
these collisions. There is no a clear evidence of the quark
constituent as well. In fact, there are numerous additional
mechanisms of a particle creation that mask the presence of the
quark-gluon plasma (QGP). It appears that the QGP could be
manifested via the observation of indirect phenomena. The natural
question arises: how to identify a useful signal that would be
unambiguously associated with a certain physical process ?

The most popular methods of analysing data produced at
relativistic heavy ion collisions are the correlation
analysis~\cite{[1]}, the analysis of  missing masses~\cite{[2]}
and effective mass spectra~\cite{[3]}, the interference method of
identical particles~\cite{[4]}. We recall that results obtained
within those methods are sensitive to assumptions made upon the
background of measurements and mechanisms included into a
corresponding model consideration. As was mentioned above, the
larger is the excitation energy, the larger is a number of various
mechanisms of the creation that should be taken into account.

As an alternative approach, one could develop a method that should
be independent on the background contribution. For instance, there
are attempts to use the maximum entropy  principle~\cite{2},
Fourier transform~\cite{22} and even by even analysis~\cite{[7]}.
Thus, a formulation of a criteria for a selection of meaningful
signals is indeed a topical objective of the relativistic heavy
ion collisions physics. The major aim of this paper is to suggest
a method that does not depend on the background information and
relies only upon the fundamental symmetries of the composite
system.

Our approach is based on Random Matrix Theory \cite{M91} that was
originally introduced to explain the statistical fluctuations of
neutron resonances in compound nuclei \cite{P65} (see also
Ref.\onlinecite{Brody}). The theory assumes that the Hamiltonian
belongs to an ensemble of random matrices that are consistent with
the fundamental symmetries of the system. In particular, since the
nuclear interaction preserves time-reversal symmetries, the
relevant ensemble is the Gaussian Orthogonal Ensemble (GOE). When
the time-reversal symmetry is broken one can apply the Gaussian
Unitary Ensemble (GUE). The GOE and GUE correspond to ensembles of
real symmetric matrices and of Hermitian matrices, respectively.
Besides these general symmetry considerations, there is no need in
other properties of the system under consideration.

Bohigas {\it et al} \cite{Boh} conjectured that RMT describes the
statistical fluctuations of a quantum systems whose classical
dynamics is chaotic. Quantum spectra of such systems manifest a
strong repulsion (anticrossing) between quantum levels, while in
non-chaotic (regular) systems crossings are a dominant feature of
spectra (see, e.g., \cite{H94}). In turn, the crossings are
observed when there is no mixing between states that are
characterized by different good quantum numbers, while the
anticrossings signal about a strong mixing due to a perturbation
brought about by either external or internal sources. Nowadays,
RMT has become a standard tool for analysing the fluctuations in
nuclei, quantum dots and many other systems (see for a review, for
example, Ref.\onlinecite{Gur}). The success of RMT is determined
by the study the statistical laws governing fluctuations having
very different origins. Regarding the relativistic heavy ion
collision data the study of fluctuation properties of the momentum
distribution of emitted particles could provide an information
about i)possible errors in measurements and ii)kinematical and
dynamical correlations of the composite system.

Let us consider the discrete spectrum $\{E_i\}, i=1,...,N$ of a
d-dimensional quantum system (d is a number of degrees of
freedom). A separation of fluctuations of a quantum spectrum can
be based on the analysis of the density of states below some
threshold $E$ \beq \label{d} S(E)=\sum_{i=1}^N\delta(E-E_i) \, .
\eeq We can define a staircase function \beq \label{s}
N(E)=\int_{-\infty}^ES(E')dE'=\sum_{i=1}^N\theta (E-E'), \eeq
giving the number of points on the energy axis which are below or
equal to E. Here \beq \label{t} \theta (x) \,=\, \left \{
\begin{array}{l}
0\qquad \,for \,\,\, x < 0 \\
1\qquad \,for \,\,\, x > 1 \\
\end{array}
\right. \eeq We separate $N(E)$ in a smooth part $\zeta(E)$ and
the reminder that will define the fluctuating part $N_{\rm fl}(E)$
\beq \label{s1} N(E)=\zeta(E)+N_{\rm fl}(E) \eeq The smooth part
$\zeta(E)$ can be determined either from semiclassical arguments
or using a polynomial or spline interpolation for the staircase
function.

To study fluctuations we have to get rid of the smooth part. The
usual procedure is to "unfold" the original spectrum $\{E_i\}$
through the mapping $E \rightarrow x$ \beq \label{m}
x_i=\zeta(E_i), \qquad i=1,...,N \eeq Now we can define spacings
$s_i=x_{i+1}-x_i$ between two adjacent points and collect them in
a histogram. The effect of mapping is that the sequence $\{x_i\}$
has on the average a constant mean spacing (or a constant
density), irrespective of the particular form of the function
$\zeta(E)$ \cite{boh2}. To characterize fluctuations one deals
with different correlation functions \cite{M91}. In this paper we
will use only a correlation function related to spacing
distribution between adjacent levels. Below, we follow a simple
heuristic argument due to Wigner \cite{Wi} that illustrates the
presence or absence of level repulsion in an energy spectrum.

For a random sequence, the probability that the level will be in
the small interval $[x_0+s,x_0+s+ds]$ is independent of whether or
not there is a level at $x_0$. Given a level at $x_0$, let the
probability that the next level be in $[x_0+s,x_0+s+ds]$ be
$p(s)ds$. Then for $p(s)$, the nearest-neighbor spacing
distribution, we have \beq p(s)ds=p(1\in ds|0\in s)p(0\in s) \eeq
Here, $p(n\in s)$ is a probability that the interval of length $s$
contains $n$ levels and $p(n\in ds|m\in s)$ is the conditional
probability that the interval of length $ds$ contains $n$ levels,
when that of length $s$ contains $m$ levels. One has $p(0\in
s)=\int_s^{\infty}p(s')ds'$, the probability that the spacing is
larger than $s$.  The term $p(1\in ds|0\in s)=\mu(s)ds$ $[\mu(s)$
is the density of spacings $s]$, depends explicitly on the
choices, 1 and 0, of the discrete variables $n,m$. As a result,
one obtains $p(s)=\mu(s)\int_s^{\infty}p(s')ds'$ which can be
solved to give \beq p(s)=\mu(s) \exp(-\int_0^s \mu(s')ds') \eeq
The function $p(s)$ and its first moment are normalized to unity,
\beq \int_0^s p(s)ds=1, \qquad \int_0^s sp(s)ds=1. \eeq For a
linear repulsion $\mu(s)=\pi s/2$ one obtains the Wigner surmise,
\beq p(s)=\frac{\pi}{2}s \exp(-\frac{\pi}{4}s^2), \qquad s\geq 0
\eeq For a constant value $\mu(s)=1$ one obtains the Poisson
distribution \beq p(s)=\exp^{-s}, \qquad s\geq 0 \eeq

As discussed above, when quantum numbers of levels are well
defined, one should expect for the spacings the Poisson type
distribution, while a Wigner type distribution occurs due to
either internal or external perturbations that destroy these
quantum numbers. In fact, one of the sources of external
perturbations can be attributed to the uncertainty in the
determination of the momentum distribution of emitted particles in
relativistic heavy ion collisions. We make a conjecture that the
above discussed ideas of the RMT are applicable  to the momentum
distribution as well. We assume that the momentum distribution may
be associated with eigenstates (quantum levels) of a composite
system. The difference between energy and momentum is inessential
for pions (see below), while we assume that the proton mass should
not affect significantly the correlation function.

Another possibilities are the association of the momentum
distribution to the spectrum of scattering matrix, or density
matrix, which can equally be the object of statistical analysis.
Note also, that here we are dealing with the momentum distribution
in the target rest frame only, postponing its comparison to that
in the center of mass frame, which is more natural for description
of interaction. Therefore, we simply replace in
Eqs.(\ref{d})-(\ref{m}) the variable $E$ by the variable $|p|$ and
construct the corresponding correlation function $p(s)$.

To test the utility and the validity of the proposal we use the
experimental data that have been obtained from the 2-m propane
bubble chamber of LHE, JINR~\cite{[8],[9]}. The chamber, placed in
a magnetic field of 1.5 T, was exposed to beams of light
relativistic nuclei at the Dubna Synchrophasotron. Practically all
secondaries, emitted at a 4$\pi$ total solid angle, were detected
in the chamber. All negative particles, except those identified as
electrons, were considered as $\pi^-$-mesons. The contaminations
by misidentified electrons and negative strange particles do not
exceed 5$\%$ and 1$\%$, respectively. The average minimum momentum
for pion registration is about 70 MeV/c. The protons were selected
by a statistical method applied to all positive particles with a
momentum of $|p|>500$ MeV/c (we identified slow protons with
$|p|\le 700$ MeV/c by ionization in the chamber). In this
experiment, we had got 20407 ${}^{12}{CC}$  interactions at a
momentum of 4.2A GeV/c (for methodical details see~\cite{[9]})
contents 4226 events with more than ten tracks of charged
particles. Thus, it was known in advance the accuracy of
measurements for available range of the momentum distribution of
secondary particles. Consequently, our analysis has been done for
different range of values of the momentum distribution to
illuminate the degree of the accuracy.

On Fig. 1 the dependence $dN/d{|p|}$ as a function of the measured
momentum (0.15-7.5 GeV/c) of the secondary particles is displayed
. The numerical data $N(p)$ were approximated by the polynomial
function of the sixth order and we obtain the distribution of
various spacings $s_i$ in 2636 events satisfying the condition of
$\chi^2$ per degree of freedom less than 1.0. Momenta are well
defined in the region 0.15-1.14 GeV/c (region I, Fig. 2a), where
the minimal value of the proton momentum is 0.15 GeV/c. The
intermediate region (region II, Fig. 2b) covers the values
1.14-4.0 GeV/c. The region 4.0-7.5 GeV/c is the less reliable one
(Fig. 2c). The spacing probability nicely reproduces this tendency
depending on the region of the momentum distribution. The function
$p(s)$ has the Poisson distribution for the region I, where the
momentum distribution was defined with a high accuracy. The region
II corresponds to the intermediate situation, when the spacing
distribution lies between the Poisson and the Wigner
distributions. The less reliable region of the values has a Wigner
type distribution for the spacing probability (Fig. 2c). Indeed,
the distribution reflects a strong deviation from the regular
behavior, observed for the measurements with a high degree of the
accuracy.

Summarizing, we propose a method to analyse data obtained at
relativistic heavy ion collisions. The method does not depend on
the background of the measurements and provides a reliable
information about correlations brought about by external or
internal perturbations. In particular, we demonstrate that the
method manifests the perturbations due to the uncertainty in the
determination of the momentum distribution of secondary emitted
particles.

\section*{Acknowledgments}
 This work was partly supported by Grant No.\ BFM2002-03241
from DGI (Spain). R. G. N. gratefully acknowledges support from
the Ram\'on y Cajal programme (Spain).

\newpage

\begin{figure}[h]
 \centerline{
 \includegraphics[width=50mm,height=50mm]{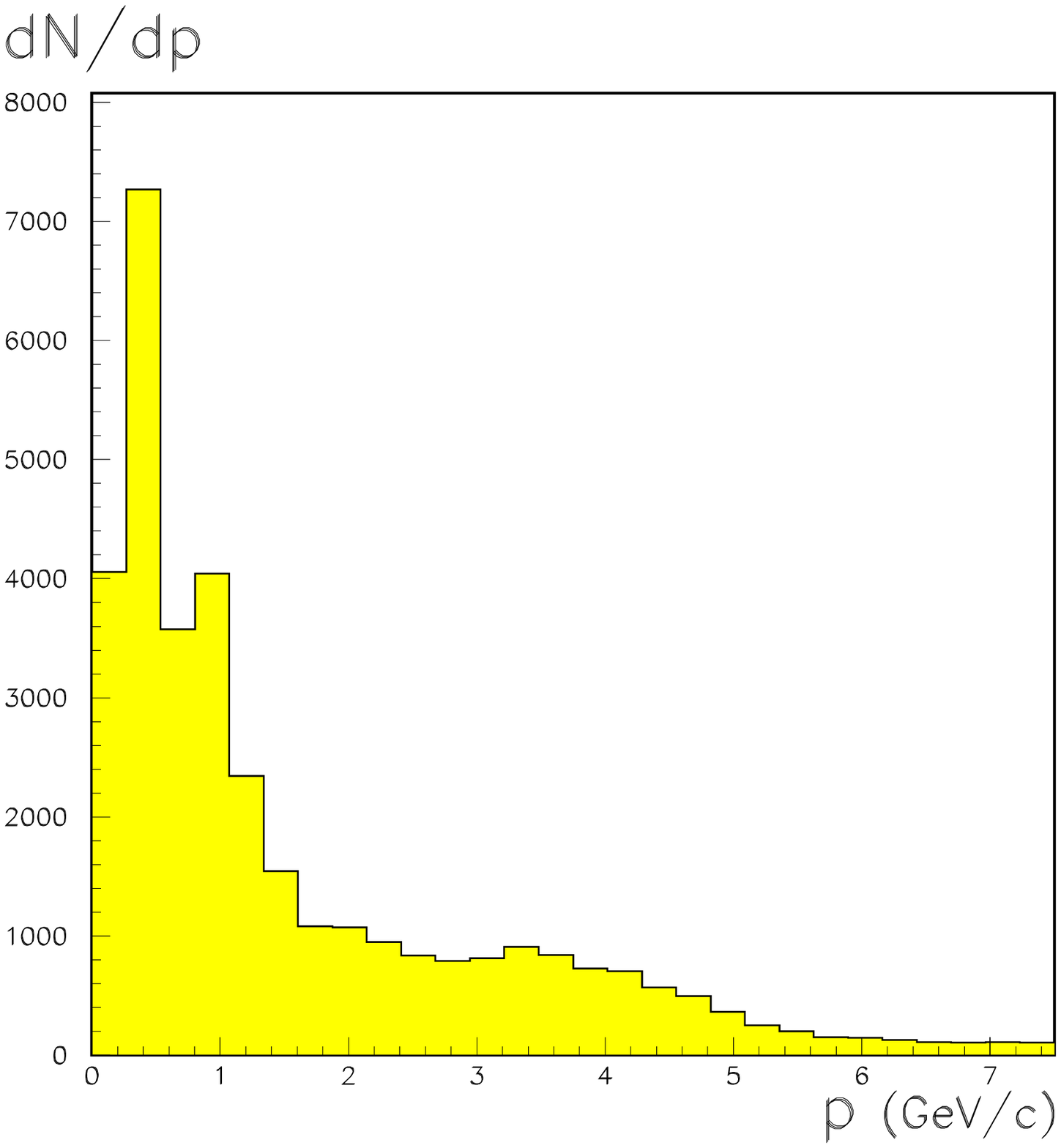}}

 {\bf Fig.1} $dN/d{|p|}$ as a function of the
           measured momentum of the secondary particles.

\end{figure}

\begin{figure}[h]
 \centerline{
 \includegraphics[width=50mm,height=50mm]{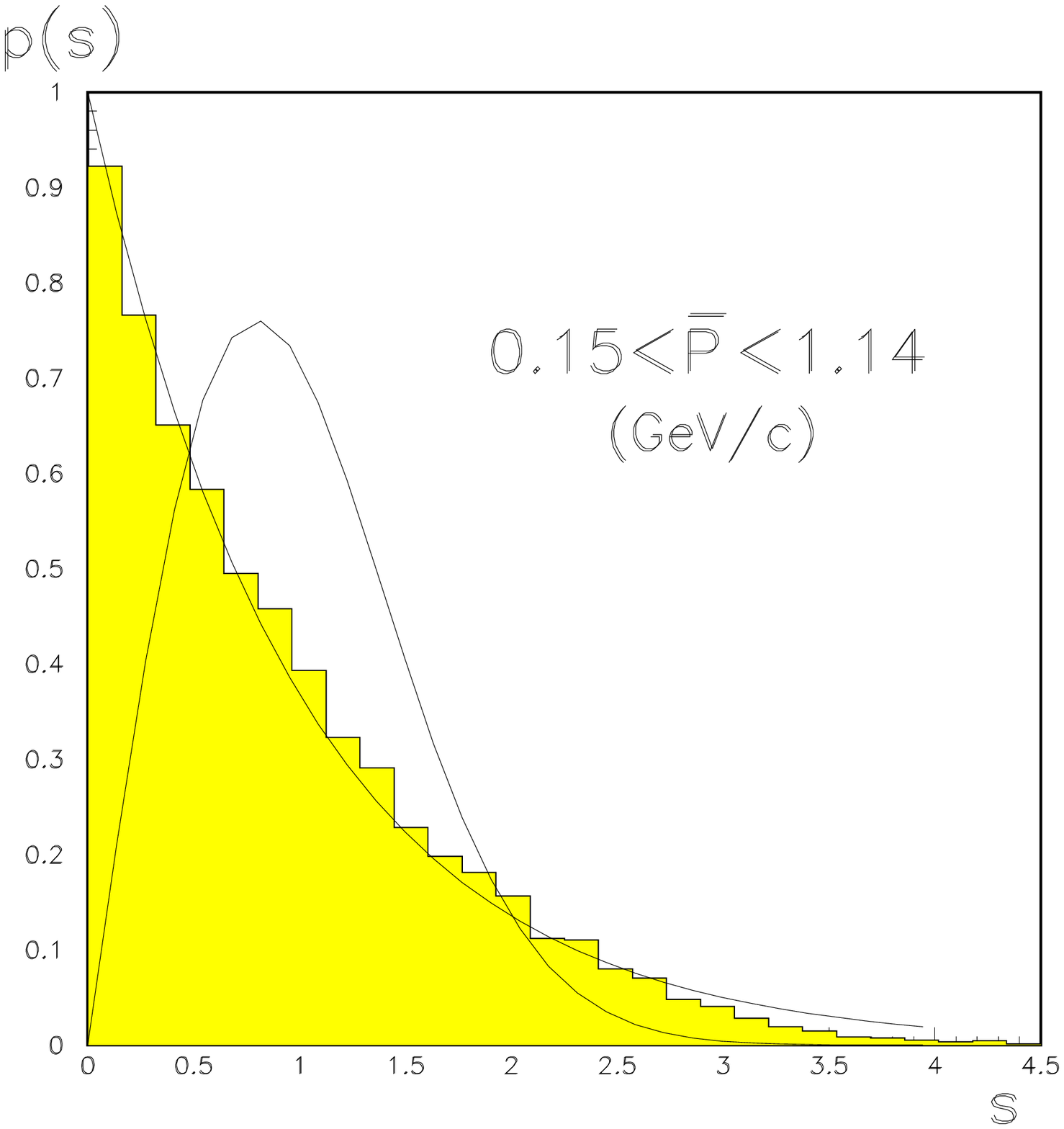}}
{\bf Fig.2 a}
\end{figure}

\begin{figure}[h]
 \centerline{
 \includegraphics[width=50mm,height=50mm]{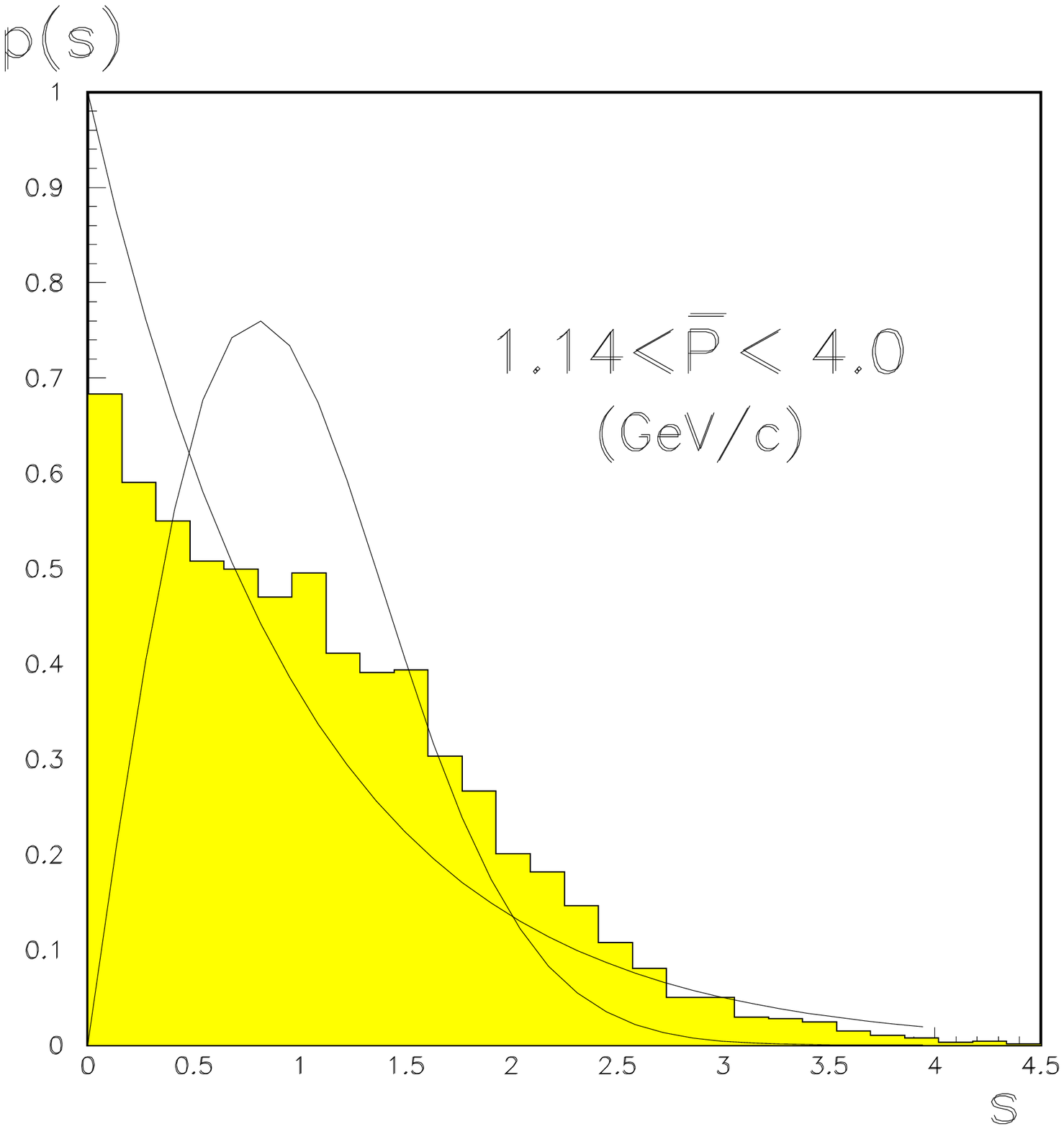}}
{\bf Fig.2 b}
\end{figure}

\begin{figure}[h]
 \centerline{
 \includegraphics[width=50mm,height=50mm]{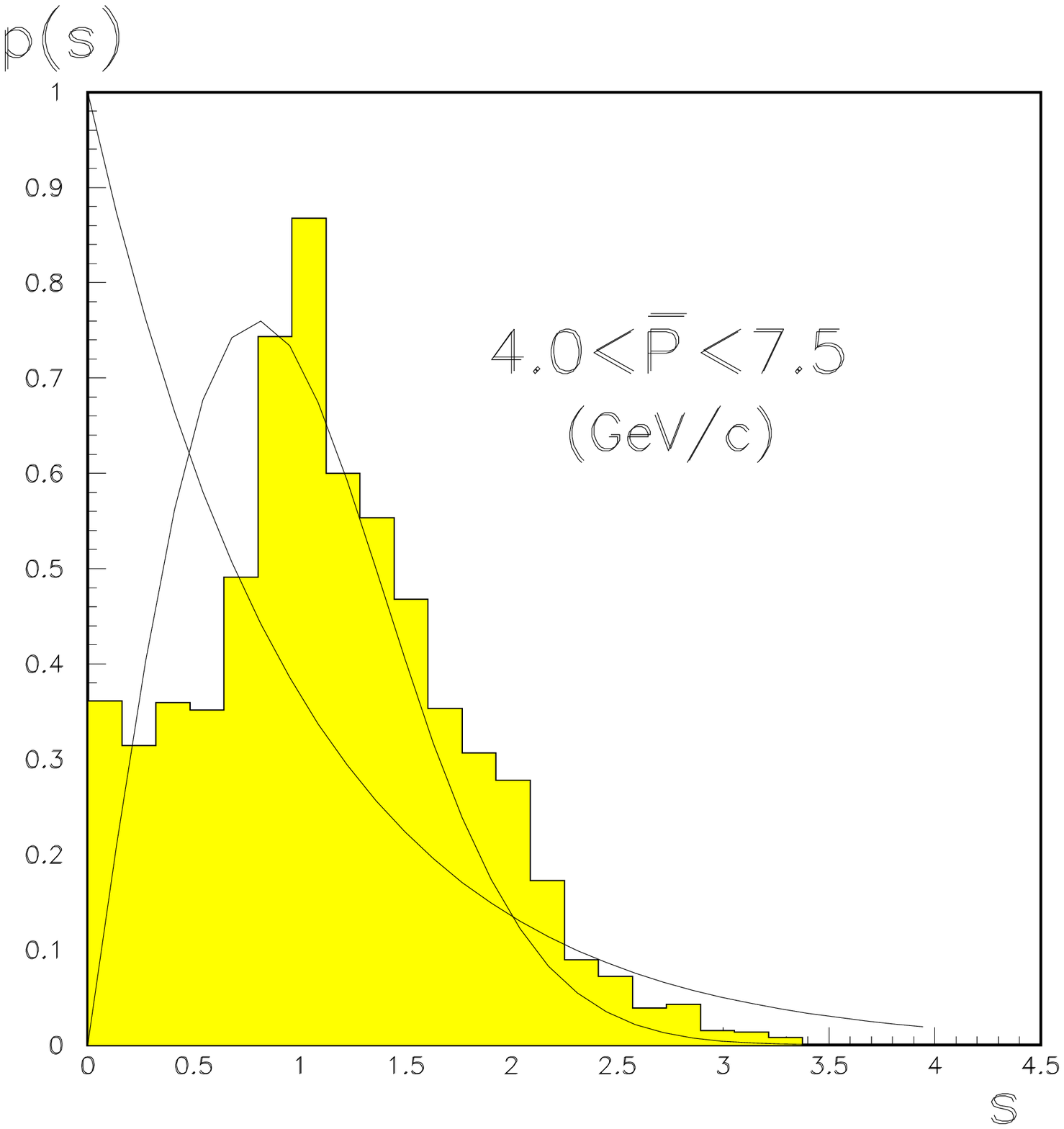}}

{\bf Fig.2 c} Nearest-neighbor spacing momentum distribution
$p(s)$
            for different  regions of measured momenta:
        a)$0.15<|p|<1.14$ GeV/c; b)$1.14<|p|<4.0$ GeV/c;
        c)$4.0<|p|<7.5$ GeV/c.
        The solid line is the Wigner-Dyson distribution and the
        dashed line is the Poisson distribution.

\end{figure}

\end{document}